\documentclass[aps,prl,twocolumn,superscriptaddress,showpacs,floatfix,nofootinbib]{revtex4-1}
\usepackage{amsmath,graphicx}

\begin{document}

\title{Determining initial-state fluctuations from flow measurements in heavy-ion collisions}

\author{Rajeev S. Bhalerao}
\affiliation{Department of Theoretical Physics, TIFR,
   Homi Bhabha Road, Colaba, Mumbai 400 005, India}
\author{Matthew Luzum}
\affiliation{
CEA, IPhT, (CNRS, URA2306), Institut de physique theorique de Saclay, F-91191
Gif-sur-Yvette, France} 
\author{Jean-Yves Ollitrault}
\affiliation{
CNRS, URA2306, IPhT, Institut de physique theorique de Saclay, F-91191
Gif-sur-Yvette, France} 
\date{\today}

\begin{abstract}
We present a number of independent flow observables that can be measured using multiparticle azimuthal correlations in heavy-ion collisions.  
Some of these observables are already well known, such as
$v_2\{2\}$ and $v_2\{4\}$, but most are new---in particular, joint correlations 
between $v_1$, $v_2$ and $v_3$.  Taken together, these measurements will allow for a more precise determination of the medium properties than is currently possible.
In particular, by taking ratios of these observables, we construct
quantities which are less sensitive to the hydrodynamic response of the
medium, and thus more directly characterize the initial-state
fluctuations of the event shape, which may constrain models for
early-time, non-equilibrium QCD dynamics. 
We present predictions for these ratios using two Monte-Carlo
models, and compare to available
data.
\end{abstract}


\maketitle

\section{Introduction}
Thermalization of the matter produced in ultrarelativistic
nucleus-nucleus collisions results in strong collective motion. 
The clearest experimental signature of collective motion is obtained
from azimuthal correlations between outgoing particles. 
It has been recently realized~\cite{Alver:2010gr}  that
fluctuations due to the internal structure of
colliding nuclei (previously studied in the context of elliptic
flow~\cite{Alver:2006wh}), followed by collective flow, naturally
generate specific patterns which are observed in these azimuthal 
correlations. 
In this Letter, we propose a number of independent flow measurements and study 
the possibility to constrain models of
initial-state fluctuations directly from these experimental data. 

\section{Flow observables}

Correlations between particles emitted in relativistic heavy-ion
collisions at large relative pseudorapidity $\Delta\eta$ are
now understood as coming from collective flow~\cite{Luzum:2010sp}. 
According to this picture, particles in a given event are emitted
independently according to some azimuthal distribution.  The most
general distribution can be written as a sum of Fourier
components, 
\begin{equation}
\label{Fourier}
\frac{dN}{d\varphi}=\frac{N}{2\pi}\left(1+2\sum_{n=1}^{\infty}
    v_n\cos(n\varphi-n\Psi_n)\right),
\end{equation}
where $v_n$ is the $n^{th}$ flow harmonic~\cite{Voloshin:1994mz} and
$\Psi_n$ the corresponding reference angle, all of which fluctuate event-by-event. 

The largest flow harmonic is elliptic flow,
$v_2$~\cite{Ollitrault:1992bk}, which has been extensively studied at
SPS~\cite{Alt:2003ab},
RHIC~\cite{Ackermann:2000tr,Adler:2003kt,Back:2004mh}, and
LHC~\cite{Aamodt:2010pa}.  
Next is triangular flow, $v_3$~\cite{Alver:2010gr}, which together
with $v_2$ is responsible for the ridge and shoulder structures observed in
two-particle correlations~\cite{Adare:2008cqb,Adams:2006tj}.
In addition, quadrangular flow, $v_4$, has been measured in correlation with
elliptic flow~\cite{Adams:2003zg,Adare:2010ux}. 
Finally, directed flow, $v_1$, can be uniquely separated~\cite{Gardim:2011qn} 
into a rapidity-odd part, which is the traditional directed
flow~\cite{Alt:2003ab,Adams:2003zg,Back:2005pc},  
and a rapidity-even part created
by initial fluctuations~\cite{Teaney:2010vd}, which has only been 
measured indirectly~\cite{Luzum:2010fb}.

In practice, one cannot exactly reconstruct the underlying 
probability distribution from the finite sample of particles
emitted in a given event. 
All known information about $v_n$ is inferred from azimuthal
correlations. 
Generally, a $k$-particle correlation is of the type
\begin{equation}
\label{defcor}
v\{n_1,n_2,\ldots,n_k\}=\left\langle \cos\left(n_1\varphi_1+\ldots+n_k\varphi_k\right)\right\rangle,
\end{equation}
where $n_1,\ldots,n_k$ are integers, $\varphi_1,\ldots,\varphi_k$ are
azimuthal angles of particles belonging to the same event, 
and angular brackets denote average over multiplets of particles and
events in a centrality class. Since the impact parameter orientation is uncontrolled, the only
measurable correlations have azimuthal symmetry: $n_1+\ldots+n_k=0$.  

In this work, we are interested in the global event shape. The average
in Eq.~(\ref{defcor}) is thus taken over all multiplets of particles.
More differential analyses (i.e., restricting one or several particles to a
specific $p_t$ interval) are left for future work. 
The average in Eq.~(\ref{defcor}) can be a {\it weighted\/} average,
where each particle is given a weight depending on pseudorapidity
and/or transverse momentum (if measured).  
Our goal here is to characterize initial-state fluctuations of the
event shape, which are approximately independent of
rapidity~\cite{Dumitru:2008wn}.  
Weights should therefore be chosen independent of (pseudo)rapidity,
which is a nonstandard choice for odd harmonics~\cite{Poskanzer:1998yz}. 
With a symmetric detector, one thus selects the rapidity-even part of $v_n$. 
We are concerned with experimental observables that can be constructed
from 
$v_1$, $v_2$ and $v_3$. 
The  study of $v_4$ and higher harmonics is more complicated due to the large
interference with  
$v_2$~\cite{Borghini:2005kd}, and is left for future work. 

Inserting Eq.~(\ref{Fourier}) into Eq.~(\ref{defcor}) gives 
\begin{equation}
\label{corflow}
v\{n_1,\ldots,n_k\}=\left\langle v_{n_1}\ldots v_{n_k}\cos(n_1\Psi_{n_1}+\ldots+n_k\Psi_{n_k})\right\rangle,
\end{equation}
where the average is now only over events.
To the extent that correlations are induced by collective flow, 
azimuthal correlations measure moments of the flow distribution.

The simplest $v_n$ measurement is the pair 
correlation~\cite{Wang:1991qh}, which corresponds to  the event-averaged root-mean-square $v_n$
\begin{equation}
\label{vn2}
v_n\{2\}\equiv \sqrt{v\{n,-n\}}\simeq \sqrt{\langle v_n^{\ 2} \rangle}.
\end{equation}
Higher-order correlations yield higher moments of the $v_n$ distribution:
\begin{equation}
\label{vn4}
v\{n,n,-n,-n\}\equiv 2v_n\{2\}^4-
v_n\{4\}^4\simeq\langle v_n^{\ 4} \rangle,
\end{equation}
where we have used the standard notation  $v_n\{4\}$ for the
4-particle cumulant~\cite{Borghini:2001vi}.

Finally, one can construct correlations involving mixed harmonics, as in previous analyses of $v_4$~\cite{Adams:2003zg}
and $v_1$~\cite{Borghini:2002vp}. 
The first non-trivial correlations between $v_1$, $v_2$ and $v_3$ are 
\begin{align}
\label{corr123}
v_{12}&\equiv v\{1,1,-2\},&
v_{13}&\equiv v\{1,1,1,-3\}, \nonumber \\
v_{23}&\equiv v\{2,2,2,-3,-3\},&
v_{123}&\equiv v\{1,2,-3\}.
\end{align}
These observables are new. Note that 
$v\{1,1,-2\}$
has been analyzed with rapidity-odd weights in harmonic
1\cite{Alt:2003ab,Adams:2003zg}, not with rapidity-even weights. 

One generally expects $v_1<v_3<v_2$. Thus correlations involving high
powers of $v_1$ are more difficult to measure.  
As explained in detail in Ref.~\cite{Bhalerao:2011ry},  
the analysis must be done in such a way as to isolate the correlation
induced by collective flow from other ``nonflow'' effects, which fall
into two categories: 
1) global momentum conservation, whose only significant contribution
is in $v\{1,-1\}$ and $v\{1,1,-2\}$. 
This effect can be suppressed by using the $p_t$-dependent weight
$w=p_t-\langle p_t^2\rangle/\langle p_t\rangle$ for at least one of
the particles in harmonic 1~\cite{Luzum:2010fb}. 
2) Short-range nonflow correlations, which can be suppressed by
putting rapidity gaps between some of the particles. 
As shown in Ref.~\cite{Bhalerao:2011ry}, all the correlations we have introduced are
likely to be measurable at the LHC. 

\section{Predictions}
\label{s:predictions}

\begin{figure}
\includegraphics[width=\linewidth]{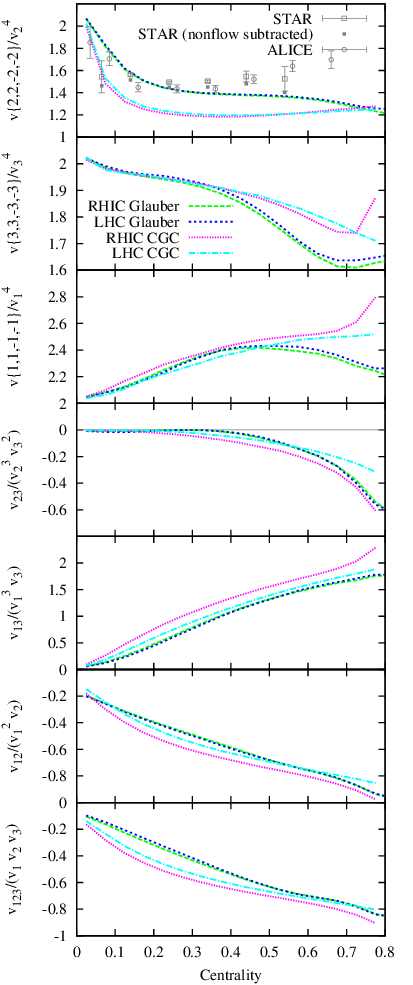}
\caption{(Color online) Predictions for ratios of various proposed measurements as a 
function of centrality (fraction of the total cross section, most
central to the left) in
Au-Au collisions at RHIC and Pb-Pb collisions at LHC, using a
Glauber- and a CGC-type model with 100 million and 20 million events, respectively.  
The factors in the denominator are 
shorthand, $v_n\equiv v_n\{2\}$.  See text for details.
}
\label{fig:ratios}
\end{figure}
The anisotropy in the distribution, Eq.~(\ref{Fourier}), has its origin in the
spatial anisotropy of the transverse density distribution at early times. 
Following Teaney and Yan~\cite{Teaney:2010vd}, we define
\begin{eqnarray}
\label{defepsilon}
\varepsilon_1 e^{i\Phi_1}&\equiv&-\frac{\{r^3 e^{i\varphi}\}}{\{r^3\}}\cr
\varepsilon_2 e^{2i\Phi_2}&\equiv&-\frac{\{r^2 e^{2i\varphi}\}}{\{r^2\}}\cr
\varepsilon_3 e^{3i\Phi_3}&\equiv&-\frac{\{r^3 e^{3i\varphi}\}}{\{r^3\}},
\end{eqnarray}
where curly brackets denote an average over the transverse plane in a
single event~\cite{Bhalerao:2006tp}, weighted by the density at midrapidity, 
and the distribution is centered in each event, $\{re^{i\varphi}\}=0$. 
In this equation, $\Phi_n$ is the minor orientation angle (corresponding, e.g., to the minor axis of the ellipse for $n$=2), and $\varepsilon_n$ the
magnitude of the respective anisotropy. 

Anisotropic flow scales like the initial anisotropy $\varepsilon_n$
and develops along $\Phi_n$. It is 
therefore natural to expect that at a given centrality, $v_n=K_n\varepsilon_n$ and
$\Psi_n=\Phi_n$, where $K_n$ is a constant that contains all information about 
the hydrodynamic response to the
initial anisotropy in harmonic $n$ --- in particular, medium properties such as the equation of state, viscosity, etc.  
These relations are not exact, but event-by-event hydrodynamic
calculations have shown that they indeed hold to a good approximation for 
$v_1$~\cite{Gardim:2011qn}, $v_2$~\cite{Holopainen:2010gz,Qiu:2011iv} 
and $v_3$~\cite{Qin:2010pf,Qiu:2011iv} --- in the majority of events, the angles
$\Psi_n\simeq\Phi_n$ are strongly correlated, while
$v_n/\varepsilon_n$ is close to a constant value for a given set of
parameters.  There are typically only small and apparently random
deviations on an event-by-event basis, making the use of this 
approximation very useful for event-averaged quantities such as those
considered here.
The validity of these relations goes beyond hydrodynamics, and they
still hold if the system is far from
equilibrium~\cite{Borghini:2010hy}. 
(On the other hand, these simple relations are not 
valid for higher harmonics such as $v_4$ and $v_5$~\cite{Qiu:2011iv},
which is why we focus here on $n\leq3$.)

Inserting these proportionality relations into Eq.~(\ref{corflow}),  we obtain 
\begin{equation}
\label{propto}
v\{n_1,\ldots,n_k\}=K_{n_1}\ldots K_{n_k}\varepsilon\{n_1,\ldots,n_k\},
\end{equation}
where we have introduced the notation
\begin{equation}
\varepsilon\{n_1,\ldots,n_k\}\equiv\left\langle\varepsilon_{n_1}\ldots\varepsilon_{n_k}
 \cos(n_1\Phi_{n_1}+\ldots+n_k\Phi_{n_k})
 \right\rangle.
\end{equation}
Thus the measured correlations are sensitive to details of the hydrodynamic 
evolution mostly through the coefficients $K_n$, and to the initial-state
 dynamics through $\varepsilon\{n_1,\ldots,n_k\}$, which only contains information about the system prior to equilibration.
 
At present, properties of the early-time system are poorly constrained
by experiment, and  
contribute the largest source of uncertainty in the extraction of 
medium properties like  
shear viscosity~\cite{Luzum:2008cw}.
However, these new proposed flow measurements will now provide an  
opportunity to significantly constrain the initial state.

In particular, one can eliminate the dependence on the proportionality
coefficients $K_n$ ---and therefore isolate initial-state effects--- by
measuring the correlations defined by 
Eq.~(\ref{defcor}), integrated over phase space, and by scaling them
appropriately: 
\begin{equation}
\label{ratios}
\frac{v\{n_1,n_2,\ldots,n_k\}}{v_{n_1}\{2\}\ldots v_{n_k}\{2\}}
=\frac{\varepsilon\{n_1,n_2,\ldots,n_k\}}{\varepsilon_{n_1}\{2\}\ldots
  \varepsilon_{n_k}\{2\}},
\end{equation}
where 
$\varepsilon_n\{2\}\equiv
\sqrt{\langle \varepsilon_n^{\ 2} \rangle}$. 
The left-hand side of Eq.~(\ref{ratios}) can be measured
experimentally, while the right-hand side depends only on early-time dynamics, and 
can be calculated using a
model of initial-state fluctuations.   Thus, although the relations \eqref{propto} are only approximate, taking these ratios minimizes sensitivity to medium 
properties and --- even though they come from correlations between soft 
particles --- they represent some of the most direct probes of initial-state dynamics available.

Eq.~(\ref{ratios}) holds if the coefficients $K_n$ are positive. With
the definitions Eq.~(\ref{defepsilon}), this always holds for $K_2$ and
$K_3$. However, $K_1$ is negative for low $p_t$
particles~\cite{Teaney:2010vd,Gardim:2011qn}. One can compensate for
this negative sign by giving a negative weight to low $p_t$ particles
in Eq.~(\ref{defcor})~\cite{Luzum:2010fb}. 

We make predictions for these ratios by computing them with two of the most common models for the initial state of a heavy-ion collision.
First is the PHOBOS Glauber Monte-Carlo~\cite{Alver:2008aq}, with binary collision 
fraction $x=0.145$ for RHIC collisions and $x=0.18$ for LHC.  The second uses the gluon 
density from a color-glass-condensate (CGC) inspired model --- the MC-KLN~\cite{Drescher:2007ax}, with rcBK unintegrated gluon 
densities~\cite{Albacete:2010ad}.
The main difference between the two models is that the eccentricity is
larger in the CGC model~\cite{Hirano:2005xf,Lappi:2006xc}.
Both models are fairly simple, with the only source of fluctuations being the nucleonic structure of nuclei.  In reality other sources of fluctuations could be important, and future study will be needed to fully understand the constraints imposed on the initial dynamics by these measurements.

Fig.~\ref{fig:ratios} displays predictions for all of the scaled correlations in Au-Au 
collisions at 200~GeV per nucleon pair and Pb-Pb collisions at
2.76~TeV per nucleon pair. 

The top three panels show 
 $v\{n,n,-n,-n\}/v_n\{2\}^4$= 
$\langle v_n^{\ 4}\rangle/\langle v_n^{\ 2}\rangle^2$.
For Gaussian fluctuations~\cite{Voloshin:2007pc}, 
the $n$=1 and 3 ratios are equal to 2 (i.e., $v_n\{4\}$=0), 
and likewise for $n$=2 in central collisions. 
However, this is expected only in the limit of a large system. 
A more detailed analysis~\cite{Bhalerao:2011} 
shows that, e.g.,  
$v_3\{4\}$ should be smaller than $v_3\{2\}$ only by a factor $\sim$2
in mid-central collisions, in agreement with these results. 
Note that wherever the ratio is greater than 2, the fourth cumulant $v_n\{4\}$ is undefined.
The top panel also shows
existing data from STAR~\cite{Adams:2004bi} and 
ALICE~\cite{Aamodt:2010pa}.  Neither measurement includes a rapidity
gap, and thus may contain nonflow correlations (see the discussion in Ref.~\cite{Bhalerao:2011ry}).
For STAR $v_2\{2\}$, we use both the raw data, and the value with an
estimated correction for nonflow effects~\cite{Ollitrault:2009ie}. 
The data seem to favor larger relative fluctuations than are contained in the MC-KLN model used here.

The bottom four panels display scaled mixed correlations, 
indicating non-trivial correlations between $\Psi_1$,  $\Psi_2$ and
$\Psi_3$.   
The scaled correlation $v_{23}$ indicates a negligible correlation between
$\Psi_2$ and $\Psi_3$  
up to $40\%$ centrality in the Glauber 
model~\cite{Alver:2010gr,Nagle:2010zk}, while the CGC model predicts
a small anticorrelation.
In contrast, $\Psi_1$ has both a 
strong correlation with $\Psi_3$~\cite{Staig:2010pn} (positive
$v_{13}$) and  a (weaker) anticorrelation with 
$\Psi_2$~\cite{Teaney:2010vd} (negative $v_{12}$), though this decreases for central collisions. 
The dependence on impact parameter can be attributed to 
the intrinsic eccentricity of the nuclear overlap
region~\cite{Bhalerao:2011}. 
The strong positive correlation between $\Psi_1$ and $\Psi_3$ explains why 
$v_{12}$ and $v_{123}$~\cite{Teaney:2010vd} have the same sign and behave similarly.

\section{Conclusion}
We have proposed a new set of independent flow observables in heavy-ion collisions which can be combined to tightly constrain theoretical models.  In particular, certain ratios are constructed which are largely determined only by the initial state, and thus directly measure properties of the early-time system.  We have presented predictions for these ratios using two common Monte-Carlo models, and compared to existing data.
\begin{acknowledgments}
We thank Raimond Snellings, Wei Li and Bolek Wyslouch for discussions. 
This work is funded by ``Agence Nationale de la Recherche'' under grant
ANR-08-BLAN-0093-01 and by CEFIPRA under project 4404-2.
\end{acknowledgments}

\end{document}